\documentclass[a4paper]{jpconf}
\usepackage{graphicx}
\usepackage{amssymb} 

\begin{document}
\title{Primordial deuterium abundance at $z_{\rm abs}=2.504$ towards Q1009$+$2956}

\author{E O Zavarygin$^{1}$, J K Webb$^{1,2}$, S Riemer-S{\o}rensen$^{3,4}$ and V Dumont$^{5,1}$}

\address{$^1$ School of Physics, University of New South Wales, Sydney, NSW 2052, Australia}
\address{$^2$ DAMTP, Centre for Mathematical Sciences, Wilberforce Road, Cambridge, CB3 0WA, UK}
\address{$^3$ Institute of Theoretical Astrophysics, The University of Oslo, Oslo NO-0316, Norway}
\address{$^4$ ARC Centre of Excellence for All-sky Astrophysics (CAASTRO), NSW 2016, Australia}
\address{$^5$ Department of Physics, University of California, Berkeley, California 94720-7300, USA}

\ead{e.zavarygin@gmail.com}

\begin{abstract}
The $z_{\rm abs} = 2.504$ Lyman Limit absorption system (LLS) towards Q1009$+$2956 has previously been used to estimate the primordial deuterium abundance. Since the initial measurement by Burles $\&$ Tytler \cite{Burles1998b}, this quasar has been observed extensively with the Keck telescope, providing a substantial increase in signal-to-noise (from 60 to 147 at continuum level of Ly-$\alpha$ at $z_{\rm abs}=2.504$). We explore a set of different models for the absorption system and find that the deuterium feature is contaminated by Ly-$\alpha$ absorption from a low column density H\,{\sc i} cloud. This significantly limits precision to which one can estimate the D/H ratio in this LLS. Our final result for  this system D/H\,$=2.48^{+0.41}_{-0.35} \times 10^{-5}$ has the same relative uncertainty of 17\% as the previous estimate by~\cite{Burles1998b} despite the far higher signal-to-noise of our dataset. A weighted mean of 13 D/H measurements available in the literature (including our result) gives the primordial deuterium abundance of $({\rm D/H})_{\rm p} = (2.545 \pm 0.025)\times10^{-5}$ and the corresponding baryon density of the Universe of $\Omega_{\rm b}h^2 = 0.02174 \pm 0.00025$ marginally inconsistent with the 2015 Planck CMB data. \end{abstract}

\section{Introduction}

Most of the elements in the Universe were produced during its late evolution in stars, supernovae and cosmic rays, whilst light elements such as hydrogen, helium, lithium and their isotopes are primordial, produced during the Big Bang nucleosynthesis epoch (BBN), the first few minutes after the Big Bang (e.\,g. \cite{Weinberg2008}). The BBN yields of the light elements abundances depend on physical conditions at that time, allowing us to explore physics in the early Universe. 

For the standard BBN scenario, the relative abundance of the light species is unambiguously determined by a single parameter: the baryon-to-photon ratio $\eta$. Given its sensitivity to and monotonic dependence on $\eta$, the primordial abundance of deuterium (D/H ratio) is generally accepted as the best ``baryometer'' amongst all the primordial elements. 

The most reliable measurements of the primordial deuterium abundance come from absorption in quasar spectra by metal-poor gas clouds on the lines of sight (e.\,g. \cite{Burles1998b,Burles1998a}). Atoms of neutral hydrogen (H\,{\sc i}) and deuterium (D\,{\sc i}) in gas clouds intersecting the line of sight absorb the flux from a quasar, due to Lyman series transitions. The values of the ionisation potential of H\,{\sc i} and D\,{\sc i} are almost identical ($\Delta E=0.0037$\,eV \cite{Morton2003}). The corresponding ionization correction should be less than $0.1\%$ \cite{Cooke2016a} making D/H~$=$~D\,{\sc i}/H\,{\sc i} a reasonable approximation.

Here we present the main results of a new analysis of the previously analysed metal-poor LLS at $z_{\rm abs}=2.504$ towards the quasar Q1009$+$2956. Since the previous analysis by \cite{Burles1998b}, this quasar has been observed extensively with the Keck telescope, and the data quality available is now substantially improved. The full version of the analysis will be available at \cite{Zavarygin2017}.

\section{Data}
\label{sec:dataset}

The dataset used in this work consists of 22 individual exposures observed with the High Resolution Echelle Spectrometer (HIRES \cite{Vogt1994}) mounted on the Keck telescope.  The initial data reduction was done using the MAKEE software designed by Tom Barlow.  Echelle orders were combined and continuum-fitted using UVES\_popler software \cite{MurphyUVES2016}.  Then, using new custom-made code, the one-dimensional normalised spectra for each of the 22 exposures were checked (and corrected) for relative velocity offsets.  The 22 datasets were then co-added into four distinct one-dimensional spectra (with signal-to-noise of 57, 47, 78 and 27 near Lyman-$\alpha$ of the LLS), determined by the instrumental settings used during the observations: the spectral resolution (resolving power $R=37000$ and 49000) and binning in the dispersion direction (unbinned and $2\times$binned). In the modelling process, these four spectra were fitted simultaneously as described below.

\section{Analysis}

\subsection{Fitting}

The absorption features were fitted with Voigt profiles using {\sc vpfit}~10.2 \cite{Carswell2014}. The following assumptions were made. Redshifts of H\,{\sc i} and D\,{\sc i} are tied in each component. The ratio of D/H is assumed to be equal in all components which is a good approximation at low metallicities \cite{Fields2001, Sembach2004}, as is the case here ([Si/H]~$= -2.5$ \cite{Burles1998b}). All column densities are allowed to vary freely in the fitting procedure and can reduce to zero if a component at some particular velocity is not required by the data itself. A more stringent constraint on D/H can be obtained by solving for {\it summed} H\,{\sc i} and D\,{\sc i} column densities for the main components of the LLS. A velocity shift free-parameter is included in the model taking into account any possible residual velocity off-sets between the spectra. A zero level is fitted for transitions with saturated lines in order to correct for any possible imperfect night-sky subtraction.

\subsection{Transitions used}
\label{sec:transitions}

The entire H\,{\sc i} Lyman series of the $z_{\rm abs}=2.504$ LLS are covered by our dataset. Out of them, we select those which are not severely contaminated by absorption lines from other redshifts: Lyman~$\alpha$, $\beta$, $\gamma$, 6, 13, 14, 21-24. Due to the relatively low H\,{\sc i}, column density and because of line blending, the deuterium absorption appears most prominently in D\,{\sc i} Lyman~$\alpha$, as Fig.~\ref{fig:lya} and \ref{fig:model_profile} illustrate. In models using metal species, the following transitions were also used: C\,{\sc ii} 1334, C\,{\sc iii} 977, C\,{\sc iv} 1548 and 1550, Si\,{\sc iv} 1393 and 1402. We do not see absorption by other species commonly detected in higher H\,{\sc i} column density absorption systems such as O\,{\sc i}, Fe\,{\sc ii}, Mg\,{\sc ii}.

\subsection{Blending of the Lyman $\alpha$ line}

\begin{figure}[t]
\center{\includegraphics[width=0.95\linewidth]{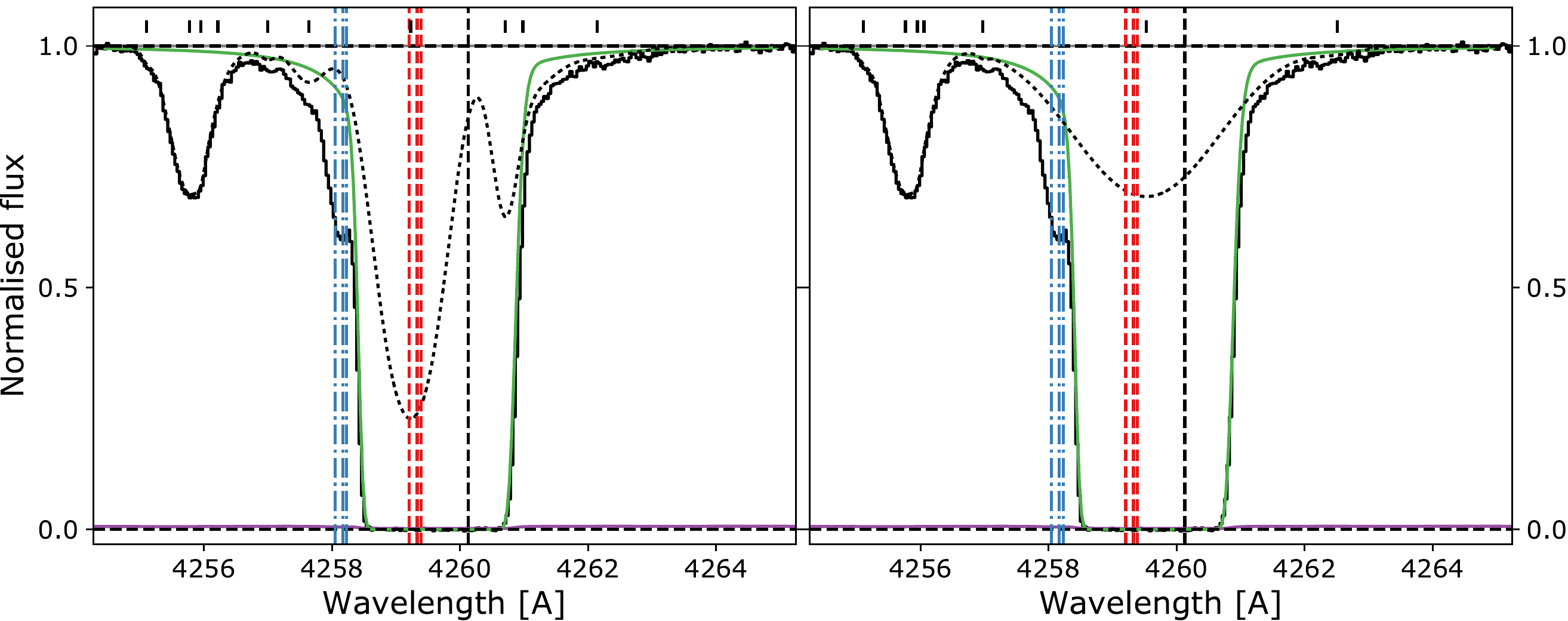}}
\caption{The Ly\,$\alpha$ transition of the $z_{\rm abs}=2.504$ LLS. The spectrum illustrated (black histogram) was obtained by a variance-weighted co-addition of the data from all 22 exposures. It is for illustration only - see Section \ref{sec:dataset}. The continuous green line is formed from the three H\,{\sc i} components (vertical red dashed lines) in models 6a (left panel) and 6b (right panel) that give rise to the D\,{\sc i} absorption feature (vertical blue dash-dotted lines), plus one moderate column density H\,{\sc i} component (vertical black dashed line) required  to account for the additional absorption redwards of the three main components.  The $1\sigma$ error array is shown by the purple line near $y=0$. The grey (very narrow) shaded area near $y=1$ corresponds to the $\pm1\sigma$ uncertainty in the continuum. The absorption due to additional low-column density blends is illustrated by the black dotted line. The positions of the blends are indicated with vertical ticks. A single broad ($b\sim105$~km/s) blend in model 6b (right panel) replaces four relatively narrow blends in model 6a (left panel).}
\label{fig:lya}
\end{figure}

Fitting the selected Lyman series transitions reveals that there is residual absorption in both wings of H\,{\sc i} Lyman\,$\alpha$ (see Fig.~\ref{fig:lya}). This absorption appears to be fairly symmetric with respect to the center of the LLS. We adopt two blending models: multiple relatively narrow ($\sim10-40$~km/s) blends on either side of the H\,{\sc i} Lyman\,$\alpha$ and a single broad (large $b$ parameter, $\sim105$~km/s) line sitting on top of H\,{\sc i} Lyman~$\alpha$. 

\subsection{Different models}
\label{sec:models}

We explore a set of models to test the sensitivity of the inferred D/H to the underlying kinematic structure of the LSS. For each blending model, discussed above, we consider six kinematic models, varying the number of absorbing components and the transitions used:\\
\indent 1) one-component model, fit to Lyman series only;\\
\indent 2) two-component model, fit to Lyman series only;\\
\indent 3) three-component model, fit to Lyman series only;\\
\indent 4) two-component model, C\,{\sc ii} used;\\
\indent 5) three-component model, C\,{\sc ii} used;\\
\indent 6) three-component model, C\,{\sc ii}, C\,{\sc iii}, C\,{\sc iv} and Si\,{\sc iv} used.\\

We use subscripts ``a'' and ``b'' to indicated models using multiple relatively narrow blends and a single broad line to model wings of H\,{\sc i} Lyman~$\alpha$: models 1a$-$6a and 1b$-$6b, respectively. To avoid degeneracy, when fitting Lyman series only (models 1$-$3), we tie the main H\,{\sc i} and D\,{\sc i} components thermally. Thermal broadening is supported by fitting to metal transitions only and to all the selected transitions making this assumption reasonable \cite{Burles1998b,Zavarygin2017}. When metal transitions are used (models 4$-$6), we solve for both temperature and $b_{\rm turb}$ parameter. When all the metals are fitted simultaneously, three velocity components are required.

\subsection{Fitting results}

\begin{figure}[h]
\center{\includegraphics[width=0.9\linewidth]{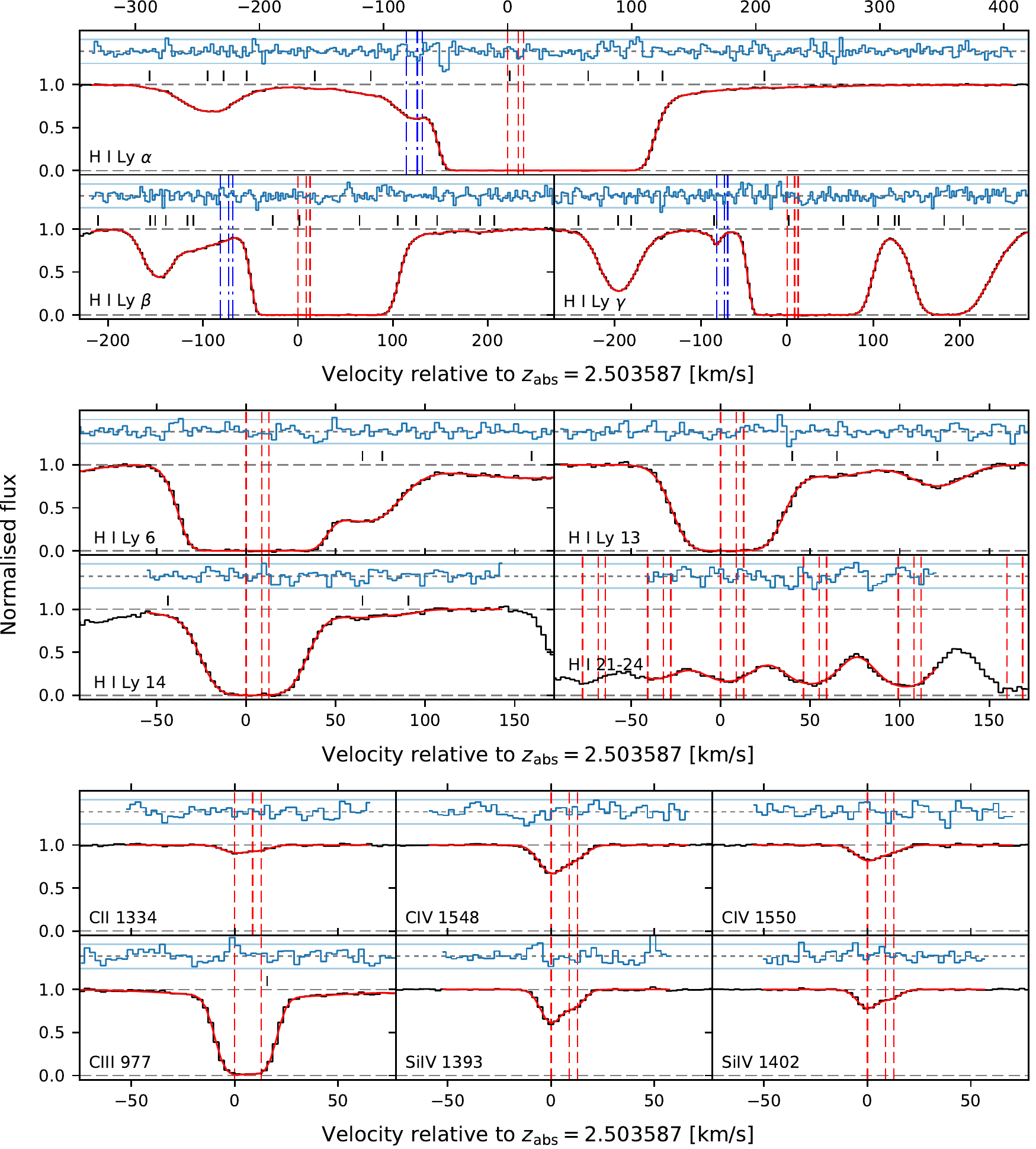}}
\caption{The overall model profile corresponding to model 6a (red solid line) for the transitions used in the analysis (indicated in the bottom left corner of each panel) over-plotted with the co-added spectrum combined from the whole dataset of 22 exposures (black histogram). This spectrum is only used for illustration, fitting was done to the four co-added spectra described in Section~\ref{sec:dataset}. The normalised residuals are illustrated above the spectra, with horizontal blue lines indicating the $\pm 2\sigma$ range. The zero point for the x-axis, in velocity units, corresponds to the bluemost (i.e. lowest redshift) component in model 6a. The red dashed vertical lines indicate the position of the three main H\,{\sc i} components. The blue dash-dotted vertical lines in the top three panels indicate the corresponding D\,{\sc i} lines. The shorter vertical lines above each fit show the position of the additional blends included in the model.
}
\label{fig:model_profile}
\end{figure}

\begin{figure}[t]
\center{\includegraphics[width=0.6\linewidth]{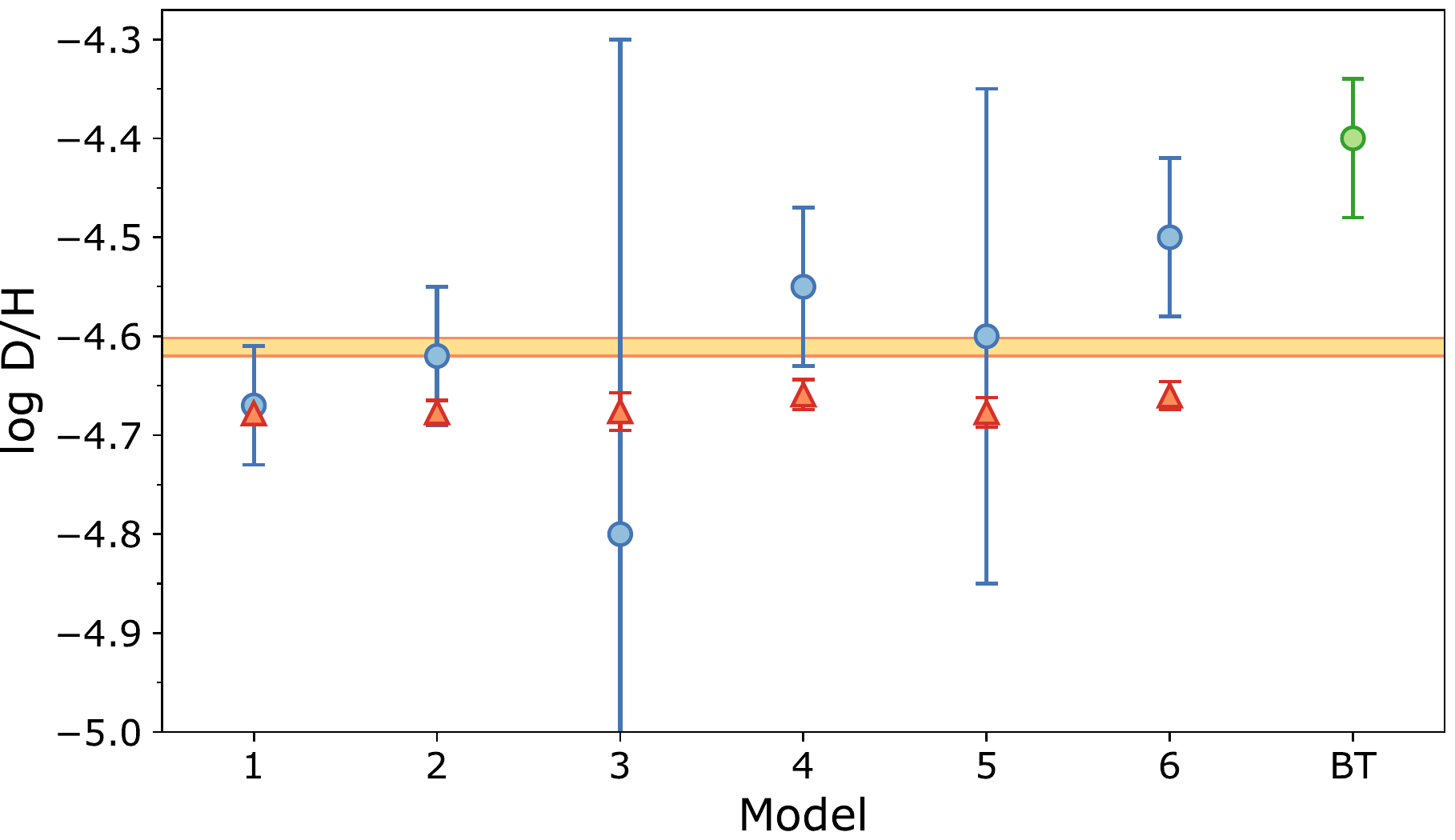}}
\caption{D/H values with their 1$\sigma$ confidence intervals (y-axis) for all considered models (indicated on the x-axis; numeration is given in Section~\ref{sec:models}). The blue points and the orange triangles correspond to models 1a$-$6a and 1b$-$6b, respectively. The green point at far right is the estimate by \cite{Burles1998b}. The CMB prediction by \cite{Coc2015} is shown with the horizontal band.}
\label{fig:d2h_results}
\end{figure}

Fig.~\ref{fig:model_profile} illustrates the overall profile for model 6a over-plotted with the spectrum co-added from the whole dataset. Fig.~\ref{fig:d2h_results} illustrates the final D/H values for all 12 models considered.  The original estimate by~\cite{Burles1998b} and the CMB prediction by~\cite{Coc2015} are also illustrated. 

Using multiple narrow blends to model the residual absorption in the wings of H\,{\sc i} Lyman\,$\alpha$ significantly degrades the precision since the position of one blend coincides closely to the strongest D\,{\sc i} line, Lyman\,$\alpha$. However, we see that D/H values for models 1a$-$6a are in good agreement with each other. Model 3a is a 3-component fit to the hydrogen series only, that is, no metals have been used.  Model 5a is also a 3-component fit, using the hydrogen series plus only one metal transition, C\,{\sc ii}, which is rather weak (Fig.~\ref{fig:model_profile}).  These two models give a conspicuously larger D/H uncertainty compared to the other models (Fig.~\ref{fig:d2h_results}), primarily because the deuterium column density is poorly constrained. Using additional metal transitions for three-component models (3a, 5a, and 6a) reduces the degeneracy and provides a better constraint on D/H. Models using a broad line to model the wings of H\,{\sc i} Lyman\,$\alpha$ also give consistent D/H values among the b models. However, unlike models 1a$-$6a, a broad blend removes the severe parameter degeneracy, resulting in far smaller D/H uncertainties in models 1b$-$6b. Whilst the broad line based models disagree with the CMB prediction by~\cite{Coc2015}, we seek to derive D/H {\it independently} from other constraints, including the CMB prediction.

\subsection{Systematic error and final D/H}

Our two sets of D/H models agree with each other within the errors although the broad-blend models (the ``b'' models) values fall systematically low compared to the narrow-blend models, ``a''. For our final constraint, we initially consider models 6a and 6b, since they use more transitions, and thus more information, to constrain the LLS velocity structure. The D/H values for these models are log\,D/H\,$=-4.50 \pm 0.08$ and $-4.660 \pm 0.014$, respectively.  We make a crude estimate of the systematic error associated with choice of blending model to be half the difference between these two D/H values, i.e. $\sigma_{\rm sys} ({\rm log\,D/H}) = 0.08$.  This value is really a rough estimate of the total error rather than just a systematic term, so is conservative. Since the $\sigma_{\rm sys}$ error is random in nature, we add it in quadrature to the statistical-only D/H uncertainties of models 6a and 6b. Our final D/H estimate for this LLS is a weighted average of the two values:
$$
{\rm log(D/H}) = -4.606 \pm 0.066 \qquad {\rm or} \qquad {\rm D/H} = 2.48^{+0.41}_{-0.35} \times 10^{-5}
$$
and is in excellent agreement with the CMB prediction by \cite{Coc2015} of D/H\,$ = (2.45\pm0.05)\times 10^{-5}$. 

Note, the precision of our result is 17\% which is the same as that of the previous D/H estimate by \cite{Burles1998b}, despite the far higher S/N of our dataset. However, the D/H we derive is lower compared to \cite{Burles1998b}.

\section{The primordial deuterium abundance and $\Omega_b h^2$}

We combine the measurement described in this paper with 14 other measurements of the D/H ratio using quasar spectra from the literature -- 12 listed in Table~4 of \cite{RiemerSorensen2017}, excluding a previous measurement of the LLS reported here, plus two additional measurements from \cite{Srianand2010, Cooke2017}.

To accommodate the possibility that scatter amongst these 15 measurements exceeds the nominal uncertainties, we apply the method of Least Trimmed Squares (LTS) to identify and discard outliers. An LTS trimming fraction of 0.85 is adopted, i.\,e. we discard 15\% of measurements (which in our case is obviously two measurements). The discarded measurements turn out to be those from \cite{Srianand2010,Pettini2001}. The weighted mean of the remaining 13 points provides a high precision estimate of the primordial deuterium abundance and is
\begin{equation}
({\rm D/H})_{\rm p} = (2.545 \pm 0.025)\times10^{-5}.
\label{eq:weighted_d2h}
\end{equation}
This value is in good agreement with the CMB prediction by~\cite{Coc2015}. For the 13 D/H measurements in our final (i.e. after LTS) sample, $\chi^2/\mu = 0.97$ (where $\mu=12$ is the number of degrees of freedom), so the scatter amongst these 13 points is consistent with the published errors. If we exclude the measurement presented in this paper and re-calculate the weighted mean D/H from the remaining 12 measurements, the value is only marginally different: $({\rm D/H})_{\rm p} = 2.546\times10^{-5}$, with the same uncertainty as Eq.~\ref{eq:weighted_d2h}.

We use our estimate of the weighted mean primordial deuterium abundance (Eq.~\ref{eq:weighted_d2h}) and the fitting formula for the BBN calculations (eq.~6 from~\cite{Balashev2016}) to infer the baryon density of the Universe, the statistical uncertainty, and the uncertainty from nuclear data parameters:
$$
\Omega_{\rm b}h^2 = 0.02174 \pm 0.00013_{\textsc{qso}} \pm 0.00021_{\rm nucl}.
$$
This compares to the Planck 2015 TT$+$lowP$+$lensing value \cite{Planck2015} (see their table 4) of $\Omega_{\rm b}h^2 = 0.02226 \pm 0.00023$, i.e. there is a marginal tension at the 1.6$\sigma$ level with our result. It is interesting to see that the overall uncertainty is now dominated by imprecision from nuclear data rather than the spectral quality of the high redshift quasar data.  The latter will need experimental improvement if we are to check on this marginal 1.6$\sigma$ discrepancy.

Further details about this analysis can be found in~\cite{Zavarygin2017}. We are refining the fit at this stage so the final result for D/H in this LLS may alter slightly.

\section*{Acknowledgments}

The conference travel was funded via the Postgraduate Research Student Support (PRSS) Scheme of UNSW. EOZ is supported by an Australian Government Research Training Program (RTP) Scholarship. This work is based on observations collected with the Keck Observatory Archive (KOA), which is operated by the W.\,M. Keck Observatory and the NASA Exoplanet Science Institute (NExScI), under contract with the National Aeronautics and Space Administration. Part of the analysis and figures were done using publicly available python packages: matplotlib~\cite{Hunter2007}, numpy~\cite{vanderWalt2011}, astropy~\cite{Astropy2013}.
This research made use of NASA's Astrophysics Data System. EOZ is also grateful to GitHub for the Student Developer Pack.

\section*{References}
\bibliography{mybibliography}

\end{document}